\documentclass{article}
\usepackage{emulateapj}
\usepackage{float,epsfig,psfig}

\newcommand{\cxo}{{\sl Chandra}}
\newcommand{\xmm}{{\sl XMM-Newton}}
\newcommand{\ros}{{\sl ROSAT}}
\newcommand{\ngc}{{NGC~4214}}
\newcommand{\msun}{M$_{\odot}$}
\newcommand{\ergl}{ergs~s$^{-1}$}

\newcommand{\cxou}{CXOU~J121538.2+361921}\newcommand{\cxous}{CXOU~J121538}

\newcommand{\ergcms}{ergs~cm$^{-2}$~s$^{-1}$}

\newcommand{\hst}{{\sl Hubble}}
\newcommand{\etal}{et al.}

\slugcomment{Submitted to Astrophysical Journal}

\begin{document}

\title{Discovery of a 3.6-hr Eclipsing Luminous X-Ray Binary in the
Galaxy NGC 4214}

\author{
Kajal K. Ghosh\altaffilmark{1},
Saul Rappaport\altaffilmark{2},
Allyn F. Tennant\altaffilmark{3},
Douglas A. Swartz\altaffilmark{1},
David Pooley\altaffilmark{4,5}, and
N. Madhusudhan\altaffilmark{2}
}
\altaffiltext{1}{Universities Space Research Association,
       NASA Marshall Space Flight Center, VP62, Huntsville, AL, 35805}
\altaffiltext{2}{Dept. of Physics and Kavli
Institute for Astrophysics and Space Research,
       Massachusetts Institute of Technology, 77
Massachusetts Ave., Cambridge, MA, 02139}
\altaffiltext{3}{Space Science Department,
       NASA Marshall Space Flight Center, VP62, Huntsville, AL, 35805}
\altaffiltext{4}{Astronomy Department, University of California Berkeley,
       601 Campbell Hall, Berkeley, CA, 94720}
\altaffiltext{4}{\cxo\ Fellow}

\begin{abstract}
We report the discovery of an eclipsing X-ray binary with a 3.62-hr
period within 24\arcsec\ of the center of the dwarf starburst galaxy
NGC 4214.  The orbital period places interesting constraints on the
nature of the binary, and allows for a few very different interpretations.
The most likely possibility is that the source lies within NGC 4214
and has an X-ray luminosity, $L_x$, of up to $7 \times 10^{38}$~\ergl.
In this case the binary may well be comprised of a 
naked He-burning donor star with a neutron-star accretor, 
though a stellar-mass black-hole accretor cannot be completely excluded.  
There is no obvious evidence for a strong stellar wind in the X-ray orbital
light curve that would be expected from a massive He star; thus, the
mass of the He star should be $\lesssim 3-4$~\msun.  
If correct, this would represent a new class of very luminous 
X-ray binary~--~perhaps related to Cyg X-3.
Other less likely possibilities include a conventional low-mass X-ray 
binary that somehow manages to produce such a high X-ray luminosity 
and is apparently persistent over an interval of years; or  a
foreground AM Her binary of much lower luminosity that fortuitously 
lies in the direction of NGC 4214.
Any model for this system must accommodate the lack of an
optical counterpart down to a limiting magnitude of 22.6 in the visible.

\end{abstract}


\section{Introduction}

With the advent of the sub-arcsec X-ray imaging capability of the \cxo\
    X-ray Observatory (Weisskopf \etal\ 2003), observations of galaxies
    out to the Virgo Cluster routinely detect tens
to hundreds of X-ray sources above
    detection limits of $\sim 10^{37\pm1}$~\ergl\
(Fabbiano \& White 2006).
In analogy with our own Milky Way, the majority of these bright sources are
    likely X-ray binaries.
The precise nature of many of these objects is
usually difficult to quantify because
    of a lack of high-quality X-ray spectra and light curves.

In a few instances, periodic dips are apparent in
the observed X-ray light curves
of individual sources.
By interpreting these dips as eclipses by a companion star, the light curves
    can be used to constrain the system orbital parameters and even the mass of
    the compact object if suitable additional information is available such as
an estimate of the companion star mass via
spectral and luminosity-class typing (see, e.g.,
Weisskopf et al. 2004; Pietsch \etal\ 2004; Pooley \& Rappaport 2005;
Fabbiano et al. 2006).

\ngc\ is a dwarf starburst galaxy, morphological type IAB(s)m, located 3.5~Mpc
distant (1\arcsec$=$17~pc). The brightest X-ray source in \ngc, \cxou\
(hereafter, \cxous), shows distinct
behavior indicative of an eclipse at a period of 3.62~hrs. This periodicity is
visible in each of 5 X-ray observations taken over a 10-yr timespan though no
more than two cycles are evident in any single observation. The X-ray light
curves and spectra of the source are presented in \S~2.
Upper limits from a search for an optical counterpart are given in \S~3.
An analysis of the X-ray source and its companion based on the
observational evidence is derived in \S~4.  In
\S~5 we discuss the interpretation of the
observations and the implications for the nature
of the binary.

\section{X-ray Timing and Spectra}

Table~1 lists the X-ray observations of \ngc\
that include the variable source \cxous.
Previous analysis of the initial \cxo\ and the
\xmm\ observations are reported in
    Hartwell \etal\ (2004).
They list source \cxous\ as source~11 at
    $\alpha=12^h15^m38.^s25$,
$\delta=36^{\circ}19\arcmin21.\arcsec4$
(J2000 coordinates).
It is by far the brightest point source among the
20 discrete sources they detect,
    with a 0.3--8.0~keV flux of 3.0$\times$10$^{-13}$~\ergcms.

\begin{center}
\begin{tabular}{llcc}
\multicolumn{4}{c}{{\sc Table 1}} \\
\multicolumn{4}{c}{X-ray observations of \cxou} \\
\hline \hline
    \multicolumn{1}{c}{Date of}    &  Instrument &  Exposure & Counts$^a$ \\
    \multicolumn{1}{c}{observation}&             & (ks)      & ($\pm$~error) \\
\hline
1994-12-10   &ROSAT/HRI       & 42.6 & 91$\pm$12  \\ 
2001-10-16   &Chandra/ACIS-S  & 26.4 & 943$\pm$31 \\ 
2001-11-22   &XMM-Newton/PN   & 16.7 & 453$\pm$36 \\ 
2004-04-03   &Chandra/ACIS-S  & 27.2 & 201$\pm$16 \\ 
2004-07-30   &Chandra/ACIS-S  & 28.6 & 843$\pm$30 \\ 
\hline
\multicolumn{4}{l}{$^a$background-subtracted in 0.5--8.0 keV energy band}
\end{tabular}
\end{center}

\begin{figure*}[t]
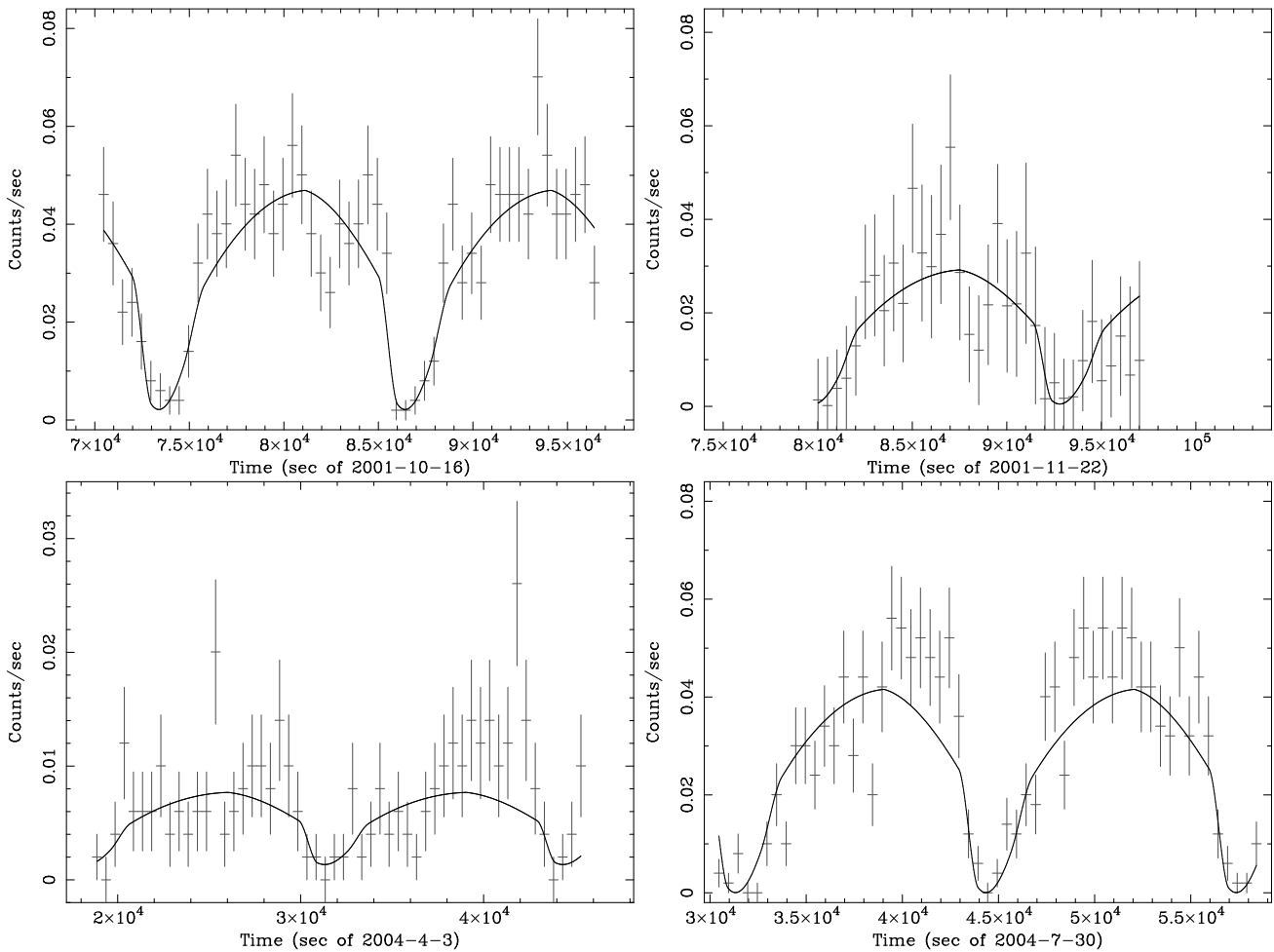

\begin{center}
\includegraphics[angle=-90,width=0.97\columnwidth]{f1a.eps} 
\includegraphics[angle=-90,width=0.97\columnwidth]{f1b.eps}
\includegraphics[angle=-90,width=0.97\columnwidth]{f1c.eps}
\includegraphics[angle=-90,width=0.97\columnwidth]{f1d.eps}
\figcaption{
Background-subtracted \cxo\ and \xmm\ X-ray light curves of \cxous.
The solid curve traces the best-fit 7-knot spline fit to the folded light
curve (see text).
\label{fig:alllc}}
\end{center}
\end{figure*}

\begin{center}
\includegraphics[angle=-90,width=0.47\textwidth]{f2.eps} 
\figcaption{Folded light curve of the X-ray dataset made by combining the
two brightest \cxo\ observations obtained on 2001-10-16 and 2004-7-30.
The heavy curve denotes the best-fitting 7-knot spline model to this
light curve.
\label{fig:foldedlc}}
\end{center}

Our locally-developed analysis software suite,
{\tt LExtrct} (Tennant 2006), was
    used to extract and analyze
    the source and background light curves and
spectra from the \ros\ and \cxo\ data.
The \xmm\ Science Analysis System (version is
6.5.0) was used for the \xmm\ data.
The  nearest discrete source is 23.\arcsec 7 to the north and is only 10\% of
    the flux of \cxous.
Thus, source \cxous\ is easily isolated from other
discrete sources even in the \xmm\ and
    \ros\ data.
For the \cxo\ data, a circular extraction region of radius 5\arcsec\ was used
    for the source and a nearby region of area 170 square arcsecs was chosen
    for the background.
For the \xmm\ data, a circular source region of radius 15\arcsec\ and a
background of 1385 square arcsecs were used for analysis.

The 0.5--8.0~keV \cxo\ and \xmm\ light curves of \cxous\ are shown in
    Figure~\ref{fig:alllc} (the \ros\ light curve is
sparsely sampled because of the
    90~min orbit of the satellite, and is not shown here). Prominent dips
are seen in all four light curves.
Using the longest exposure data, the \cxo\ observation on 2004-7-30,
    the light curve was folded
    on trial frequencies ranging from
6.4$\times$10$^{-5}$ to 1.2$\times$10$^{-4}$~Hz
    (some noise is apparent at lower frequencies).
The resulting $\chi^2$ statistic {\sl vs.} trial
frequency was fit to a Gaussian
    distribution. The peak of the Gaussian is at 7.682(60)$\times$10$^{-5}$~Hz
    corresponding to a period of 13020(100)~s.
A repeat of this procedure for the first \cxo\ dataset, from
2001-10-16, resulted in a
    best-fit period of 12900(170)~s.

Since both datasets have the same period, within the uncertainties,
we combined them to increase the number of counts per phase bin.
The minima of the two datasets were aligned via a  $\chi^2$ fitting
procedure before they were added together. 
The folded light curve made from this combined dataset (in 64 phase
 bins, Fig.~\ref{fig:foldedlc}) was fit using a 7-knot spline with
 periodic boundary conditions to create a model light curve.
This model was then applied to all the light
curves as shown in Fig.~\ref{fig:alllc}.
For these individual exposures, only the amplitude, phase, and DC level
  of the spline model were
    allowed to vary in the model fitting.
A period of 13020(100)~s is consistent with all the observations.
However, since the uncertainty in the period is of order 1\%, phase
coherence is totally lost in 100 cycles or
approximately two weeks; much shorter
    than the time between observations.
For this analysis, no barycenter corrections were applied because
they should be at most
$\sim$30 km~s$^{-1} \times 10^4$ s or about 1 lt-sec,
which is well within our measurement
errors.

\begin{center}
\includegraphics[angle=0,width=0.47\textwidth]{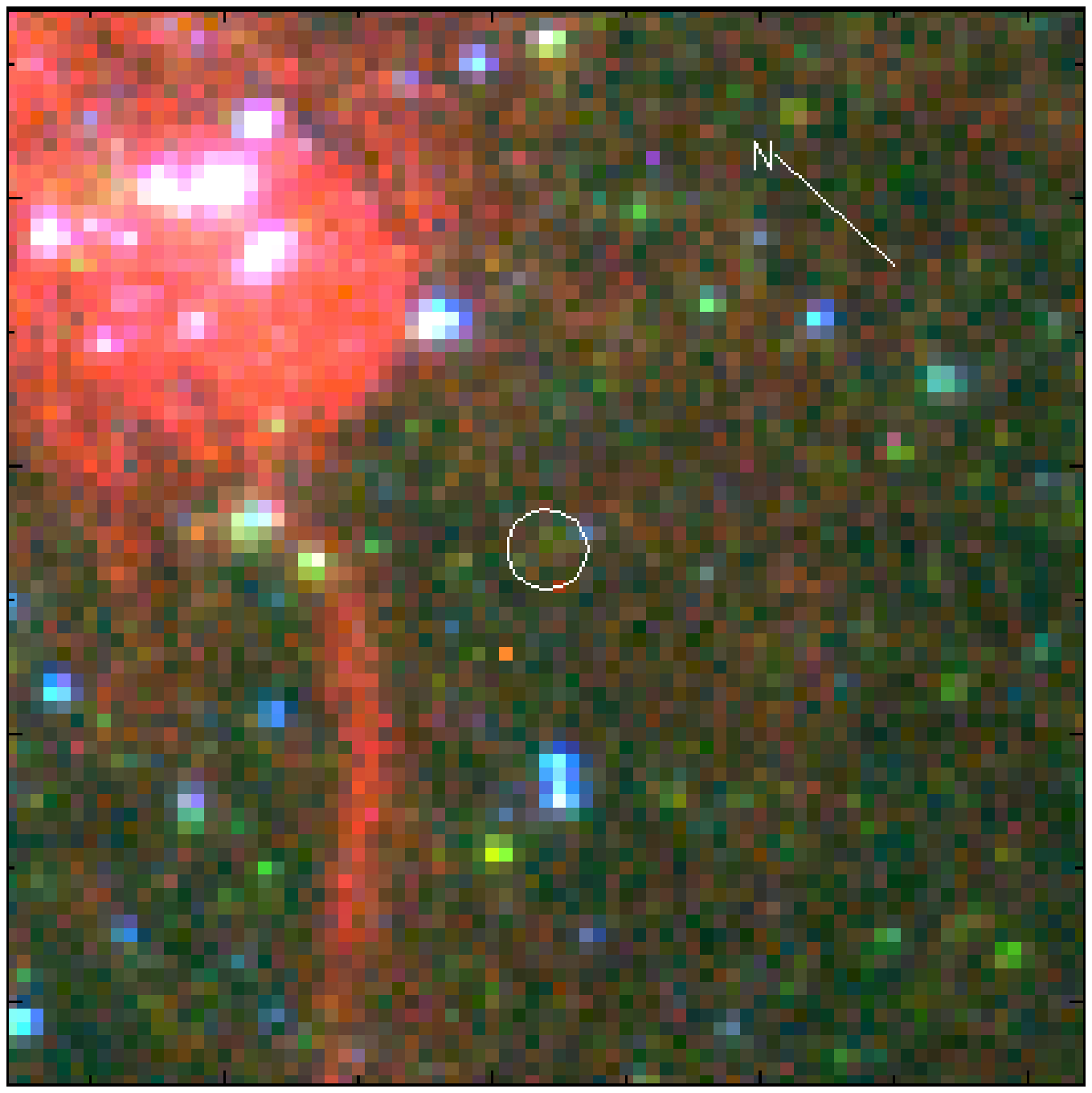} 
\figcaption{Three-color \hst\ image of an 8\arcsec $\times$8\arcsec\
region around \cxous\ in \ngc.
Data are from 1997 (MacKenty \etal\ 2000) and include
F656N (red), F555W (green), and F336W (blue) filters.
The error region for \cxous, with a 1/3$''$ radius,
is shown as a circle near the image center. North is indicated.
\label{fig:hst}}
\end{center}

Several models provide acceptable fits to the spectra of \cxous\ extracted from
    the individual \cxo\ and \xmm\ observations.
The model fit parameters are consistent with the values obtained by
 Hartwell \etal\ (2004); namely, for an absorbed power law model,
 $N_{\rm H} = (1.8^{+0.6}_{-0.7})\times 10^{21}$~cm$^{-2}$ and
 $\Gamma = 1.8^{+0.3}_{-0.2}$.
The average flux in the 0.5--8.0~keV band ranges from
    (2.0$\pm$0.1)$\times$10$^{-14}$~\ergcms\ (on 2004-04-03) to
    (3.41$\pm$0.34)$\times$10$^{-13}$~\ergcms\ (on 2001-10-16).
The flux during the bright phases on 2001-10-16 is
  (4.77$\pm$0.48)$\times$10$^{-13}$~\ergcms.
Fewer than 1000 source counts were accumulated during
any individual X-ray observation
    of \cxous\ (Table~1).
Therefore, no other important constraints on the nature of the source can
    be deduced from these spectra.

If the dips in the light curve are due to an eclipse, then the spectral shape
    during time intervals near the minimum in the light curve may differ from
    the shape at other times due to changes in
absorption or scattering of X-rays by the
    atmosphere or wind of the companion star.
To test for this possibility, we compared spectra extracted from high- 
 ($>$0.035~c~s$^{-1}$) and low-count-rate ($<$0.025~c~s$^{-1}$) 
 phases of the 2001-10-16 \cxo\ observation.
A Kolmogorov-Smirnoff test showed these spectra are consistent with being 
 drawn from the same parent distribution at the 40\% level. 
Examination of the 
 soft (0.5--2.0) and hard (2.0--8.0~keV) light curves also show
 no evidence for enhanced absorption near or away from eclipse.

In summary,
periodicity is apparent in all the observations of \cxous. We surmise the
    periodicity is due to an eclipse of the X-ray source by a companion
star and that the orbital period of the system is
    $P_{\rm orb} = 13020(100)~{\rm s} = 3.62(3)~{\rm h}$.

\section{Optical Counterpart Search} \label{sec:optical}

\hst\ observations of \ngc\ were carried out in 1997 in several filters
    as reported by MacKenty \etal\ (2000).
The dynamical center of \ngc\ (\cxo\ source~10 in Hartwell \etal\ 2004)
and \cxous\ are both located on the WPC2 camera~3 in these
observations making the registration between the optical and X-ray images
 accurate.
We estimate the radius of the X-ray error circle to be $\sim$0.\arcsec 3.
There are several optical sources within this circle in the \hst\ images
   (Figure~\ref{fig:hst}).
Taking the total light within the error circle as a conservative upper limit
    to any potential counterpart, we deduce the following observed magnitudes:
    $m_{\rm F336W}=23.6$~mag, $m_{\rm F555W}=23.7$~mag, 
 $m_{\rm F702W}=22.4$~mag,
    and $m_{\rm F814W}=22.0$~mag.
Since the different sources within the error circle contribute different
    amounts in the different bandpasses, optical colors 
 based on the total light are not meaningful.
An estimated reddening correction can be made from the fitted X-ray absorption
    column, which averages to $\sim$2$\times$10$^{21}$~cm$^{-2}$,
corresponding to
    A(V)$\sim$1.1~mag.
The resulting upper limit to the optical
counterpart to \cxous\ is $m_{\rm F555W}=22.6$~mag.
By way of comparison, we note that this is about equivalent to
   a late OV~star at the distance of \ngc.

\section{Orbital Constraints}

\subsection{Orbital Period}

In order to understand what this eclipse discovery implies, we start
by examining the constraints on the donor star set by the measured
orbital period, $P_{\rm orb} =3.62$ hr.  First we take Kepler's 3rd
law:
\begin{equation}
\frac{GM_T}{a^3} = \left(\frac{2\pi}{P_{\rm orb}}\right)^2 ~~,
\end{equation}%
where $M_T$ and $a$ are the total binary mass and orbital separation,
respectively, and utilize the relation: $R_L = r_L a = R_{\rm don}/f$,
where $R_L$ is the Roche-lobe radius of the donor
and $f$ is the fraction of the Roche lobe that is filled by the donor
star of radius $R_{\rm don}$.  Combining these, we find:
\begin{equation}
\frac{{R}^{3/2}_{\rm
don}}{f^{3/2}{M}_T^{1/2}r_L^{3/2}} \simeq 0.36
\left(\frac{P_{\rm orb}}{1\,{\rm hr}}\right)~~,
\end{equation}
where ${M}_T$ and ${R}_{\rm don}$ are in solar units.  Finally, we make
use of Eggelton's (1983) expression for $r_L$ to find:
\begin{equation}
\frac{{R}^{3/2}_{\rm don}}{{M}_{\rm don}^{1/2}} \simeq 0.12
\xi(q)f^{3/2}\left(\frac{P_{\rm orb}}{1\,{\rm hr}}\right)~~,
\end{equation}
where
\begin{equation}
\xi(q) \equiv
\frac{q^{1/2}\sqrt{1+q}}{\left[0.6q^{2/3}+\ln(1+q^{1/3})\right]^{3/2}}~~,
\end{equation}
and $q$ is the mass ratio $M_{\rm don}/M_{\rm acc}$.  Here, $M_{\rm
acc}$ is the mass of the accreting star.  Equation (3)
is analogous to the classical radius-mass relation (e.g.,
eq.\,[3] of Pooley \& Rappaport 2005; see also
the discussion in Weisskopf et al.\,2004) for
Roche-lobe filling donor
stars, but is generalized via Eggleton's (1983) expression for $r_L$
for virtually any mass ratio.  The function $\xi(q)$ differs by less
than 10\% from unity for $0.01 \lesssim q \lesssim 2$, and increases
to a value of only 1.35 as $q \rightarrow 10$.  An additional feature
of eq.\,(3) is that the factor $f$ allows for donor stars that
underfill their Roche lobes.

\begin{center}
\includegraphics[angle=0,width=0.47\textwidth]{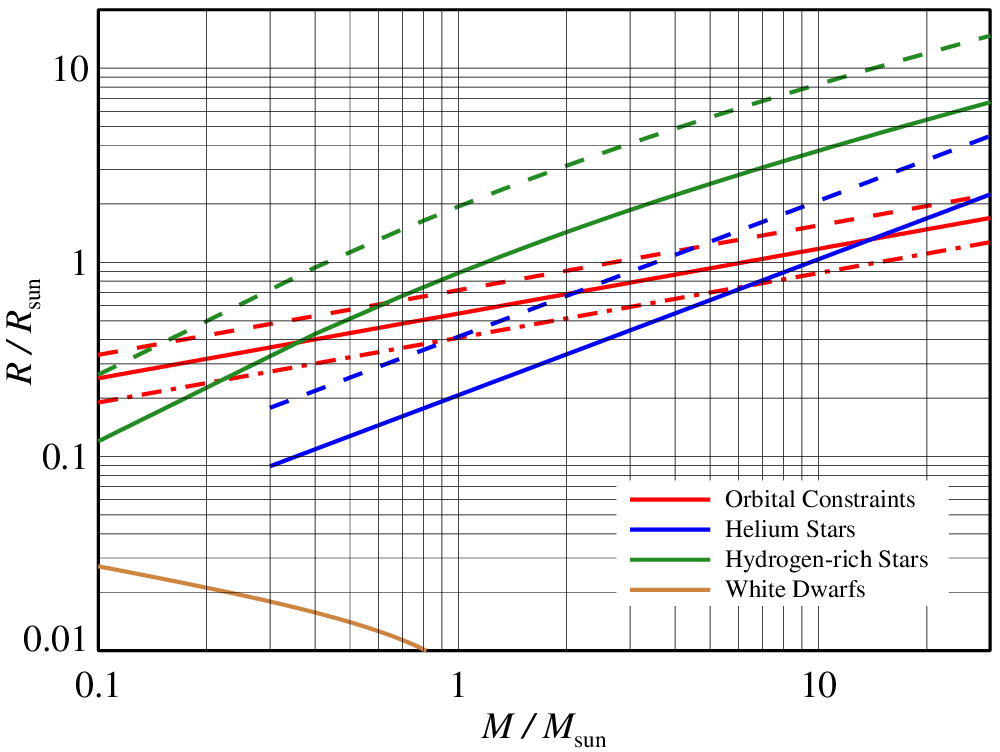} 
\figcaption{Constraints on the eclipsing source \cxous\ in \ngc.
Red curves: radius-mass
constraints set by the value of $P_{\rm orb}$; solid, dashed and dot-dashed
curves correspond to the mass ratio,
$q \equiv M_{\rm don}/M_{\rm acc}$,
and Roche-lobe filling parameter, $f$, in the following pairs
(1,1), (10,1) and (1,3/4), respectively. Green
curves: H-burning stars; solid (dashed) curve represents the zero
(terminal) age main sequence.   Blue curves: He-burning stars;
solid (dashed) curve represents the zero (approximate terminal)
age main sequence.  Orange curve: degenerate (He) dwarfs.
\label{fig:RofM}}
\end{center}

A plot of eq.\,(3), with $P_{\rm orb} = 3.62$ hr, is shown in
Figure~\ref{fig:RofM} in the radius--mass plane of the donor star.
The solid (dashed) red curve is for all $q \lesssim 2$ ($q = 10$);
both curves are for $f=1$ (i.e., a Roche-lobe filling donor). The lower
red dot-dashed curve represents the case where the donor star fills
only 3/4 of its Roche lobe, and $q=1$. The solid (dashed) green curve
is a simple $R(M)$ relation for zero age (terminal age) H-burning
main-sequence stars, while the solid (dashed) blue curve is the $R(M)$
relation for zero age (terminal age) He-burning main-sequence stars
(see, e.g., Paczy\'nski 1971; Kato \& Iben 1992; Justham \& Podsiadlowski
2006, private communication). For completeness, we show, as an orange
curve, the $R(M)$ relation for degenerate He stars.

From an inspection of Figure~\ref{fig:RofM}, 
 and the intersection of the red curves
with the green and blue curves, we can draw some basic conclusions
about the nature of the donor star.

\vspace{0.3cm}\centerline{{\em
Hydrogen-burning companion
star}}\vspace{0.3cm}

One obvious interpretation
of the nature of the eclipsing binary is a normal
main-sequence donor star with a mass of $\sim$0.4~\msun\
(the intersection of the green and red curves of Figure~\ref{fig:RofM}).
This,
coupled with an orbital period of 3.62 hr, evokes either a 
conventional low-mass X-ray binary (LMXB) in NGC 4214, or a 
cataclysmic variable binary (CV) with a white dwarf accretor in our 
own Galaxy.  In the former case, an important question would be how 
an unevolved, low-mass donor star could drive a sufficiently high 
rate of mass transfer to account for $L_x \simeq 7 \times 10^{38}$~\ergl.
Mass transfer from an unevolved 0.4~\msun\ star is induced by 
 orbital decay through gravitational
 radiation and magnetic braking
 (see, e.g., Rappaport, Verbunt, \& Joss 1983). 
The mass transfer rate would then be low
 and we estimate the corresponding X-ray luminosity would range from only
 $\sim 2 \times 10^{36}$~\ergl\ (for gravitational radiation and a
 neutron star accretor) to 
 $\sim 6 \times 10^{37}$~\ergl\ (for magnetic braking and a
 stellar-mass black hole accretor). 

In the case that \cxous\ is a CV in our own Galaxy, 
the shape of the X-ray eclipse (see, Figs.~\ref{fig:alllc} 
and~\ref{fig:foldedlc}) is reminiscent of that of AM Her systems 
(Heise \etal\ 1985).
If CVs are confined to the Galactic disk with a scale height $H\sim300$~pc,
then \cxous, in this interpretation, would be within
$\sim$1~kpc, corresponding to an X-ray luminosity of
$\sim$6$\times$10$^{31}$~\ergl, or about 1/3 of the average luminosity
for AM~Her systems (see, e.g., Ramsay \& Cropper 2003).

There are no physical solutions for extremely~--~or even
  moderately~--~evolved normal main-sequence
stars which would more 
naturally drive the high rate of mass transfer 
implied for a system 
at the distance of \ngc.

\vspace{0.3cm}\centerline{{\em Helium-burning
companion star}}\vspace{0.3cm}

Another obvious possibility for the nature of the
binary is a He-burning star in a binary with either a neutron-star or
stellar-mass black-hole accretor (see also Weisskopf et al. 2004).
The intersection of the various orbital constraint (red) curves with
the He-star (blue) curves in Fig.\,4 yield a range of possible
combinations of He-star mass and accretor mass.
For example, a $\sim$12~\msun\ He-ZAMS star could fill its Roche with an
accretor of comparable mass (i.e., $q \sim 1$), e.g., a
$\sim$12~\msun\ black hole.  Alternatively, a more massive
He-ZAMS donor of $\sim$23~\msun\ would fill its Roche lobe with
a much less massive accretor, e.g., a neutron star.  Or a $\sim$7~\msun\
He-ZAMS donor could underfill its Roche lobe (by a factor of, e.g., 3/4) and
still achieve substantial mass transfer via a stellar wind.  However, for
reasons discussed below, perhaps the most promising interpretation is
an intermediate-mass He star (e.g., $\sim$2--3~\msun) that is somewhat evolved, with approximately twice its main-sequence radius, or near 
the He terminal-age main sequence (TAMS).

\subsection{Eclipse Duration}

For the case where the X-ray eclipse is due to
the compact accretor going behind the companion
star, we can also use the eclipse duration to
learn more about the constituent masses of the
binary.  Note that these constraints would
therefore pertain to the case of the He-star
companion, but probably not to an AM Her-type eclipse.
We modify the expression for the eclipse half
angle, $\theta_{\rm ecl}$, from eq.\,(4) of
Pooley \& Rappaport (2005; see also Weisskopf et al. 2004)
to allow for arbitrary mass ratios, $q$,:
\begin{equation}
\theta_{\rm ecl} =
\cos^{-1}\left[\frac{1}{\sin 
i} \sqrt{1-r_L^2}\right]
~~,
\end{equation}
where $i$ is the orbital inclination angle, and 
\begin{equation}
r_L = \frac{0.49q^{2/3}}{\left[0.6q^{2/3}+\ln(1+q^{1/3})\right]}~~,
\end{equation}
(Eggleton 1983). A plot of $\theta_{\rm ecl}(q)$ is shown in 
Fig.~\ref{fig:eclipse} for inclination angles of $90^\circ$, 
$80^\circ$, and $70^\circ$. The horizontal
lines denote our estimates for the maximum and minimum observed 
eclipse half angles.  We have not performed any detailed
eclipse calculations for the case where a stellar wind accounts for
an appreciable portion of the X-ray modulation.  Given the limited
statistical precision associated with the X-ray
light curves, we can say only that the drop in
intensity more closely resembles a classical X-ray eclipse, i.e.,
as in a high-mass X-ray binary, than a wind-modulated light curve,
e.g., X1908+075 (Levine et al. 2004).

\begin{center}
\includegraphics[angle=0,width=0.47\textwidth]{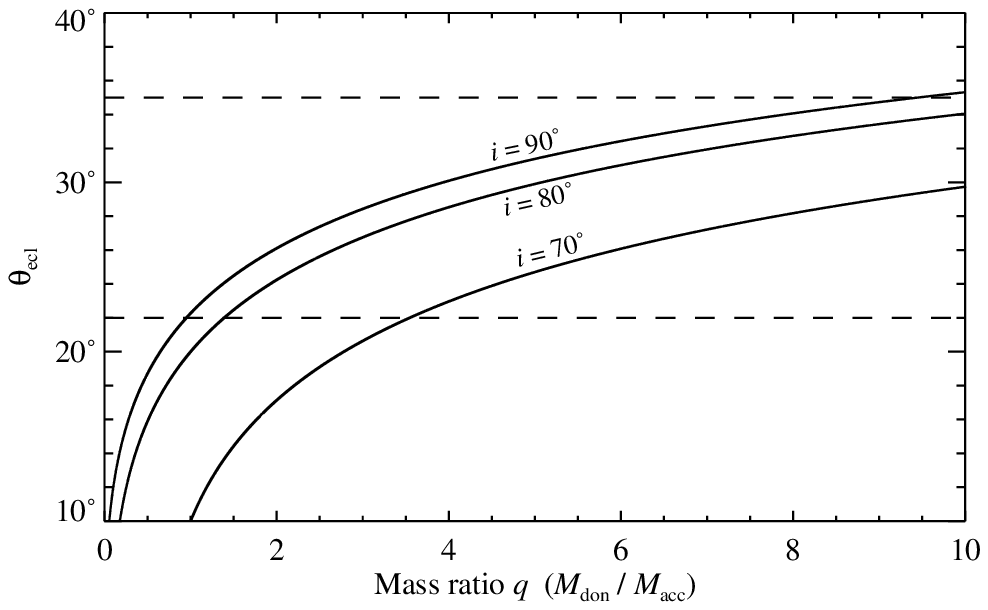} 
\figcaption{Geometric eclipse half-angle (in degrees) as a function of
the mass ratio, for several values of the inclination angle, $i$ (eq.\,[5]).  
The horizontal lines are estimates for the observed upper and lower limits on
$\theta_{\rm ecl}$.
\label{fig:eclipse}}
\end{center}

From the intersection of the two horizontal curves with the
$\theta_{\rm ecl}(q)$ curves in Fig.~\ref{fig:eclipse}, we conclude
that the mass ratio in this system is
\begin{equation}
1 \lesssim q 
  \lesssim 10 ~~.
\end{equation}
This range of $q$ values is consistent with the illustrative system
components suggested above.  In particular, unevolved He stars of mass
$\sim7-18$~\msun\ could have a black-hole accretor of
$\sim7-12$~\msun\ (intersection of solid blue with solid and
dot-dashed red curves in Fig.\,4).  Higher mass He stars (i.e.,
$\gtrsim$22~\msun) would require much less massive accretors (e.g.,
neutron stars; intersection of solid blue and dashed red curves).
Finally, somewhat evolved He stars of $\sim2-3$~\msun\ could have
neutron-star or very low-mass black hole accretors.

\section{Discussion}

Our observations of a 3.62-hr eclipse in the source
\cxous\ have yielded three distinct possibilities
for the nature of the binary.  We discuss the
pros and cons of each in turn.

\subsection{Conventional LMXB}

As discussed above, in the context of the orbital constraints, if the 
donor is a H-rich, low-mass star it cannot be very evolved.  
Thus, it is unclear how an unevolved, low-mass donor star could drive a
sufficiently high rate of mass transfer to account for values of 
$L_x$ that approach $7 \times 10^{38}$ ergs s$^{-1}$ during several different observations spanning nearly 3 years~--~especially if
the accretor were a neutron star.
Conventional black hole LMXBs,
 on the other hand, can have high luminosities but 
 only during short transient outbursts.
Systems with short orbital periods like \cxous\ 
should either not be transients or, if they 
are, should not have such prolonged intervals of high $\dot M$ (see, e.g., 
King \etal\ 1996; Kalogera \etal\ 2004; Fabbiano \etal\ 2006).  
Moreover,
if the accretor were a stellar-mass black hole then $q \ll 1$ and
 the eclipse duration would be much shorter than observed 
 (Figure~\ref{fig:eclipse}).

In discussing a possible short orbital period ULX, Liu \etal\ (2002) 
suggested beaming of the X-radiation to reduce the actual value of 
$L_x$.  Given that \cxous\ is an eclipsing system this possibility seems 
remote, though not impossible if the accretor were a rapidly spinning black 
hole, and a large kick during its formation oriented the orbital
plane perpendicular to the spin axis.

\subsection{CV/AM Her system}

For the case of an AM Her system,
the eclipse would be caused by periodic occultations of
the accreting magnetic polar cap as it rotates
around the spin axis of the white dwarf.  In this
case, we cannot infer anything further about the
mass ratio of the binary constituents, as in
Fig.~\ref{fig:eclipse}.  Thus, we focus here on the constraints
set by the HST image (Fig.~\ref{fig:hst}), and an evaluation
of the probability of finding a high Galactic
latitude ($b \simeq 78^\circ$) AM Her binary
within the $D_{25}$ area of \ngc. As discussed in
\S\ref{sec:optical}, the limiting visual magnitude
for any counterpart to \cxous\ is 22.6.  However,
at $P_{\rm orb} \simeq 3.62$ hr, we expect the
donor star in an AM Her system to be roughly of
intrinsic spectral type $M0~V-M3~V$ with visual
magnitude $8.9 \lesssim M_{\rm V} \lesssim 10.5$.
  From the optical limit, this requires the source to be
   at least 2.6 to 5.5~kpc away.
Thus, unless the hypothesized AM Her system is in
the Galactic halo, the observed constraint on its
magnitude seems difficult to reconcile with its
expected brightness.

It also seems
somewhat unusual that an AM Her system would be
found by chance so close ($\sim 25''$) to a
particular galaxy that is being observed with
\cxo\ for the purpose of detecting
luminous X-ray sources.  The $D_{25}$ ellipse region
associated with \ngc~has semi-minor and
semi-major axes of $197''$ and $255''$,
respectively.  This corresponds to a solid angle
of $\sim$44 square arcmin, or  $0.012$ square
degrees.  If there are $n_{\rm CV}$ CVs pc$^{-3}$ with
$P_{\rm orb} \gtrsim 3$ hr, then the surface
density, $S$, toward the NGP is $2 n_{\rm CV} H^3$~sr$^{-1}$, for
an assumed exponential distribution of CVs from
the Galactic plane with scale height $H$.  For
illustrative parameters of $n_{\rm CV} \sim 10^{-6}$
pc$^{-3}$ (see, e.g., from observations, Patterson 1984;
Howell \etal\ 2002; Schmidt et al. 2005; Gansicke et al. 2005;
Szkody et al. 2006; and from population models,
de Kool 1992; Kolb 1993; Han et al. 1995; and Politano,
Howell, \& Rappaport 1998) and $H \sim 300$ pc, 
we have $S \sim 55$ per steradian, or
$\sim$0.016 per square degree.  Thus, the a
priori probability of finding a CV with $P_{\rm
orb} \gtrsim 3$ hr within the 0.012 square degree
area of the \ngc\ $D_{25}$ ellipse seems quite small, i.e.,
$\sim$2$\times$10$^{-4}$.

Finally, in regard to finding a CV close to a
galaxy of particular interest, we note that a
similar circumstance has arisen in the case of CG
X-1 in the field of the Circinus galaxy
(Weisskopf et al.\,2004).  This particular
eclipsing source had an orbital period of 7.5 hr,
which already is quite long for an AM Her system.
The Circinus galaxy is at $b \simeq -3.8^\circ$
(compared to $78^\circ$ for \cxous), and
therefore the probability of finding a
chance foreground CV is higher than near the NGP.
Nonetheless, these two eclipsing sources, when
taken together, reduce the likelihood that both
are foreground CVs.

Overall, we find the hypothesis that \cxous\ is a
foreground AM Her system to be implausible.

\subsection{Naked He-burning donor star}

If we interpret the \cxous\ system as
having a naked He-burning donor star of
mass somewhere in the range of $\sim 2-20$~\msun,
the requisite mass-transfer
rate, $\dot M$, could be achievable via either
direct Roche-lobe overflow or via Bondi-Hoyle
capture (see, e.g., Bondi \& Hoyle 1944) of a
modest fraction of the stellar wind from the He
star (Dewi et al.\,2002; Justham et al.\,2006). For the
lower-mass He stars within this range, direct Roche-lobe
overflow would probably be required to supply
the mass transfer rate.  The nuclear lifetimes of He stars
in this mass range
are quite short ($\sim$3$\times$10$^5$ to 3$\times$10$^6$ yr;
Paczy\'nski 1971; Habets 1986; Langer 1989; Kato \& Iben
1992; Dewi et al.\,2002; Justham \& Podsiadlowski 2006,
private communication) so that $\dot M$ values of
$\gtrsim 10^{-7}$~\msun\ yr$^{-1}$ are easily attained.

If the donor star is indeed a He-burning star, there is an additional
constraint that the X-ray light curve neither appears
to be dominated by modulation in a dense stellar
wind, nor suffers from any obvious photoelectric
(i.e., strongly energy-dependent) absorption.
For a simple $1/r^2$ stellar wind profile, the
column density as a function of orbital phase (at
$i = 90^\circ$) has a simple analytic form:
\begin{equation}N = n_0 a \phi/\sin \phi ~~~,
\end{equation}
except for values of $\phi$ corresponding to direct geometric
eclipse (e.g., Levine et al. 2004), where $n_0$ is the wind density
at the orbit of the compact accretor, $a$ is the orbital radius, and $\phi$
is the phase angle (with $\phi = 180^\circ$ being
defined as superior conjunction).  For some
illustrative parameters, we find:
\begin{equation}N \simeq 10^{23} \dot M_{-6}
\left(\frac{a}{3\,R_\odot}\right)^{-1}
v_{1000}^{-1} ~ {\rm cm}^{-2}~~~,
\end{equation}
where the units for $\dot M$ and $v$ are
$10^{-6}$~\msun\ yr$^{-1}$ and 1000 km
s$^{-1}$, respectively.  Thus, unless the stellar
wind is essentially completely ionized, the
optical depth to soft X-rays would be enormous.
For stellar-wind density profiles that start with
zero velocity and infinite density at the stellar
surface (e.g., Lucy \& Solomon 1970; Castor,
Abbott, \& Klein 1975), the column densities
would be larger, and the discrepancy with
the observed light curve would be even greater.
Therefore, if the donor star in \cxous\ is
ultimately identified with a He star with $M \gtrsim 3-4$~\msun,
this apparent lack of a dense stellar wind will have to be
reconciled with the binary stellar model.

To quantify the constraint on the He-donor star set by the
apparent lack of a strong stellar wind, we utilize the following
expression for stellar-wind rates as a function of luminosity by
Hamann, Koesterke, \& Wessolowski (1995):
\begin{equation}
\log \dot M (M_\odot\, {\rm yr}^{-1}) = -11.95 + 1.5 \log (L/L_\odot)
~~,
\end{equation}
where $L$ is the bolometric luminosity of the He star.  More
recently, Hamann \& Koesterke (1998) and Petrovic, Langer \& van der
Hucht (2005) have suggested that the wind loss rates are lower than
given by eq.\,(10) by factors of a few. Only for luminosities
$\lesssim 10^{4}\,L_\odot$ does the donor star avoid producing a
highly attenuating stellar wind (e.g., with
$\dot M \lesssim 10^{-6}$~\msun\ yr$^{-1}$).  This luminosity,
in turn corresponds to He star masses $\lesssim 3.5$~\msun\
(Kato \& Iben 1992; Justham \& Podsiadlowski 2006, private communication).  
Our limit on the optical
counterpart of 22.6 magnitudes, is also much more consistent with
lower mass He stars (i.e., $2-3$~\msun).

\section{Summary and Conclusions}

The brightest X-ray source in the field of the
galaxy \ngc, \cxou, shows distinct
eclipse-like intensity dips at a period of 3.62~hr.
These eclipses are present in each of five X-ray
observations spanning 1994 to 2004.
Assuming that these eclipses are indicative of the
orbital period, we can rather tightly constrain the
nature of the binary system. 

We have considered the possibility that \cxous\ is a 
conventional LMXB that somehow produces a very high luminosity for a 
sustained interval of time.  The mass transfer rate that can be 
driven by a low-mass, unevolved donor star is too low to supply the 
requisite mass transfer rate, and it is unlikely that this source is, 
or should be, a transient.  

We have also evaluated quantitatively
the possibility that the \cxous\
system is a foreground CV (of the AM Her
subtype). Major difficulties with this scenario
are (i) the improbability of finding an AM Her system
aligned so closely, by chance, with a galaxy of
interest (even after taking into account the fact
that {\em Chandra} has observed a substantial
number of such galaxies with equal exposure), and
(ii) the lack of an optical counterpart brighter than
   22.6~mag, requiring the CV be a halo object.
We consider this
possibility unlikely.

The alternative is that the donor is a He-burning star.
He-donor star masses $M\gtrsim 4$~\msun\ are difficult
  to reconcile with the apparent lack of a dense stellar wind.
We believe that the most likely interpretation of this system is a
$\sim2-3$~\msun\ slightly evolved (e.g., TAMS) naked He star
transferring matter (stably) through the inner Lagrange point to a
neutron star. This type of binary has been extensively modeled by
Dewi et al.\,(2002).  If confirmed, this would open
exciting possibilities for both stellar evolution
studies (see, e.g., Dewi et al.\,2002; Justham et al. 2006) and the
interpretation of ultraluminous
X-ray sources at the lower end of their luminosity
function (see, e.g., Colbert \& Miller 2004;
Fabbiano \& White 2006). 
In the case of \cxous, the peak observed luminosity is only
 about twice the Eddington limit for accretion of He onto a neutron star.
If the accretor in such a system, i.e., with a He donor star, were a 
 stellar-mass black hole, then the Eddington limit
 would be $\sim 4 \times 10^{39}$~\ergl; well within
 the range of ultraluminous X-ray source luminosities.

Finally, we note that \cxous\ may be the first known
{\sl immediate} progenitor of a compact double neutron
star binary; i.e., beyond the second 
 common envelope phase and prior to collapse
 of the He star core.
A perhaps unverifiable prediction is that
within some $10^5-10^6$ years the evolved core of the He
star will collapse to produce a type Ib supernova explosion
(see, e.g., Podsiadlowski, Joss, \& Hsu 1992; Woosley, Langer,
\& Weaver 1995) and leave a binary radio pulsar
(if the natal kick given to the pulsar is not so large as to unbind
the system; see, e.g., Pfahl et al. 2002 and references therein;
Podsiadlowski et al. 2004).  The putative neutron star that is currently in
the system is in the process of being spun up via accretion
torques and will be the rapidly rotating pulsar member of the
binary.  It is also interesting to note that this predicted
event has actually occurred already (given the $\sim$10
Myr light travel time for the information to reach us).

\acknowledgements
We thank Stephen
Justham, Philipp Podsiadlowski, Eric Pfahl, and
Steve Howell for very helpful discussions.  KKG is
supported in part by NASA under grant NNG04GC86G issued
through the Office of Space Science.
DP gratefully acknowledges
support provided by NASA through Chandra Postdoctoral
Fellowship
grant number PF4-50035 awarded by the Chandra X-ray
Center, which
is operated by the Smithsonian Astrophysical
Observatory for NASA
under contract NAS8-03060.  SR received some
support from Chandra
Grant TM5-6003X.


\end{document}